\documentclass[aps,pra,preprint,onecolumn,superscriptaddress]{revtex4-1}
\usepackage{amsfonts}
\usepackage{amsmath}
\usepackage{amssymb}
\usepackage[colorlinks, citecolor=blue,linkcolor=red]{hyperref}
\usepackage{color}
\usepackage{float}
\usepackage{hyperref}
\usepackage{textcomp}
\usepackage{graphicx}

\begin{document}

\title{Vortex solitons in fractional nonlinear Schr\"{o}dinger equation with
the cubic-quintic nonlinearity}
\author{Pengfei Li}
\email{lpf281888@gmail.com}
\address{Department of Physics, Taiyuan Normal University, Jinzhong, 030619, China}
\address{Institute of Computational and Applied Physics, Taiyuan Normal University, Jinzhong, 030619, China}
\author{Boris A. Malomed}
\address{Department of Physical Electronics, School of Electrical Engineering, Faculty of Engineering, and Center for Light-Matter Interaction, Tel Aviv University, Tel Aviv 69978, Israel}
\author{Dumitru Mihalache}
\address{Horia Hulubei National Institute of Physics and Nuclear Engineering, Magurele, Bucharest, RO-077125, Romania}

\begin{abstract}
We address the existence and stability of vortex-soliton (VS) solutions of the
fractional nonlinear Schr\"{o}dinger equation (NLSE)\ with competing
cubic-quintic nonlinearities and the L\'{e}vy index (fractionality) taking
values $1\leq\alpha \leq 2$. Families of ring-shaped VSs with vorticities $s=1,2
$, and $3$ are constructed in a numerical form. Unlike the usual
two-dimensional NLSE (which corresponds to $\alpha =2$), in the fractional
model VSs exist above a finite threshold value of the total power, $P$. Stability
of the VS solutions is investigated for small perturbations governed by the
linearized equation, and corroborated by direct simulations. Unstable
VSs are broken up by azimuthal perturbations into
several fragments, whose number is determined by the fastest growing
eigenmode of small perturbations. The stability region, defined in terms of $P$,
expands with the increase of $\alpha$  from $1$ up to $2$ for all $s=1$, $2$, and $3$,
except for steep shrinkage for $s=2$ in the interval of $1\leq\alpha\leq 1.3$.
\end{abstract}

\maketitle

\section{Introduction}

\label{Sec I}

It is well known that optical beams propagating in conservative and
dissipative self-focusing media readily build spatial optical solitons,
which are a subject of great interest to fundamental and applied studies in
diverse settings \cite{Trillo}-\cite{Rosanov2}. In particular, a class of 2D
vortex solitons (VSs) feature a bright ring shape with an embedded rotating
screw phase dislocation, see early original works \cite{early}-\cite{early6}
and early reviews \cite{Review1}-\cite{RaduVolkov}, as well as recent ones
\cite{Review2}-\cite{Review6}. VSs usually represent excited states of the
corresponding nonlinear systems. In particular, VSs in the simplest
two-dimensional (2D) model, based on the nonlinear Schr\"{o}dinger equation
(NLSE) with cubic self-focusing \cite{Kruglov-PLA1985}, may be considered as
excited states of the fundamental mode known as Townes solitons \cite%
{TownesSoliton}. In\ models with basic self-attractive nonlinearities, such
as those represented by quadratic and cubic terms, self-trapped vortex
states are subject to azimuthal symmetry-breaking instabilities which split
them into sets of fragments, that may be fundamental solitons \cite%
{Ele-Lett33-608,PRL79-2450}. Therefore, stability is a crucially important
issue in studies of VSs. Several methods for the stabilization of VSs in
nonlinear media without the help of external potentials were suggested, such
as the use of competing quadratic-cubic \cite{QC1}-\cite{QC3} and
cubic-quintic (CQ)\ \cite{CQ1}-\cite{CQ12} nonlinearities, or,
alternatively, nonlocal self-interaction \cite{Nonlocal1}-\cite{Nonlocal7}.
In this connection, it is relevant to mention that exact solutions for
stable 1D solitons of the NLSE with the CQ nonlinearity were first obtained
in early work \cite{Bulgaria}.

Recently, the Schr\"{o}dinger equation was extended to fractional
dimensions, starting from works by Y. Hu and N. Laskin in Ref. \cite%
{FSE1a,FSE1b}. The fractional Schr\"{o}dinger equation (FSE) was introduced
as a quantum model \cite{FSE2}, in which Feynman path integrals over
Brownian trajectories lead to the standard Schr\"{o}dinger equation, while
path integrals over \textquotedblleft skipping" L\'{e}vy trajectories lead
to the FSE \cite{FSE3}. Although implications of such theories are still a
matter of debate \cite{FSE4,FSE5}, experimental schemes have been proposed
for their realization in condensed-matter settings and optics \cite%
{FSE11,FSE12}. In particular, optical cavities offer an appropriate ground
to explore intriguing properties of the FSE. Propagation of beams has been
intensively studied in the framework of FSE, producing effects such as
zigzag trajectories \cite{FSE13}, diffraction-free propagation \cite{FSE14}-%
\cite{FSE18}, beam splitting \cite{FSE19}, periodically oscillating
evolution of Gaussian modes \cite{FSE20}, beam-propagation management \cite%
{FSE21}, optical Bloch oscillations and Zener tunneling \cite{FSE22},
resonant mode conversions and Rabi oscillations \cite{FSE23}, Anderson
localization and delocalization \cite{FSE24}, and $\mathcal{PT}$-symmetric
optical modes \cite{FSE25}.

A natural extension of the analysis is to add nonlinearity to the FSE \cite%
{NLFSE2,NLFSE3}. In particular, the 1D Schr\"{o}dinger equation with the
cubic term subject to singular spatial modulation may emulate fractional
dimension $0<D<1$ \cite{Olga}. Recent works have demonstrated that the
fractional NLSE supports a variety of fractional spatial-soliton solutions
\cite{NLFSE4}: \textquotedblleft accessible solitons" \cite{NLFSE5}-\cite%
{NLFSE7}, double-hump and fundamental solitons in $\mathcal{PT}$-symmetric
potentials \cite{NLFSE8,NLFSE9}, bulk and surface gap solitons in $\mathcal{%
PT}$-symmetric photonic lattices \cite{NLFSE10}-\cite{NLFSE11a}, vortex
solitons in $\mathcal{PT}$-symmetric azimuthal potentials \cite{NLFSE11b},
2D self-trapped modes \cite{NLFSE12a}, spontaneous symmetry breaking in a
dual-core system \cite{we}, and dissipative solitons in the fractional
complex-Ginzburg-Landau equation \cite{Yingji}. Further, the composition
relation between nonlinear Bloch waves and gap solitons supported by
lattices \cite{NLFSE14,NLFSE12b}, solitons under the action of nonlinearity
subject to spatially-periodic modulation \cite{NLFSE15}, dissipative surface
solitons\cite{NLFSE15b}, as well as discrete solitons \cite{NLFSE15d}, have
also been addressed. Recently, off-site- and on-site-centered VSs in $%
\mathcal{PT}$-symmetric photonic lattices have been investigated in Ref.
\cite{NLFSE13}. However, prediction of spatial VS solutions in free space
(without the use of external potentials) in the framework of fractional NLSE
is still an open problem. This issue is addressed in the present work.

The paper is organized as follows. The model is introduced in Sec. \ref{Sec
II}, which is followed by the analysis of the stability and dynamics of
vortex-soliton states in Sec. \ref{Sec III}. The paper is concluded by Sec. %
\ref{Sec IV}.

\section{The model}

\label{Sec II}

We start the analysis by considering the beam propagation along the $z$-axis
in a nonlinear isotropic medium with the CQ nonlinear correction to
refractive index, $n_{\mathrm{nonlin}}(I)=n_{2}I-n_{4}I^{2}$, where $I$ is
the light intensity, while $n_{2}$ and $n_{4}$ are coefficients accounting
for, respectively, cubic self-focusing and quintic defocusing. The
respective fractional NLSE is%
\begin{equation}
2ik_{0}\frac{\partial A}{\partial z}-\left( -\nabla _{\perp }^{2}\right)
^{\alpha /2}A+\frac{2k_{0}^{2}}{n_{0}}n_{\mathrm{nonlin}}A=0,  \label{NLFSE1}
\end{equation}%
where $A(z,x)$ is the local amplitude of the optical field, the intensity
being $I\equiv |A|^{2}$, $k_{0}=2\pi n_{0}/\lambda $ is\ the wavenumber,
with background refractive index $n_{0}$ and optical wavelength $\lambda $,
while $-(-\nabla _{\perp }^{2})^{\alpha /2}=-(-\partial ^{2}/\partial
x^{2}-\partial ^{2}/\partial y^{2})^{\alpha /2}$ is the
fractional-diffraction operator with the corresponding L\'{e}vy index $%
\alpha $ belonging to interval $1<\alpha \leq 2$. Equation (\ref{NLFSE1})
with $\alpha \leq 1$ and cubic-only self-focusing gives rise to the wave
collapse, which destabilizes solitons; however, the quintic self-defocusing
term suppresses the instability and, in particular, makes it possible to
construct stable VSs at $\alpha =1$, as shown below.

The balance between the fractional diffraction and nonlinearity makes it
possible to build spatial solitons. By means of rescaling, $\Psi (\zeta ,\xi
)=\sqrt{n_{4}/n_{2}}A(z,x)$, $\zeta =\left( k_{0}n_{2}^{2}/n_{0}n_{4}\right)
z$, and $\left( \xi ,\eta \right) =\left(
2k_{0}^{2}n_{2}^{2}/n_{0}n_{4}\right) ^{1/\alpha }\left( x,y\right) $, Eq. (%
\ref{NLFSE1}) is cast in the normalized CQ form,

\begin{equation}
i\frac{\partial \Psi }{\partial \zeta }-\left( -\nabla _{\perp }^{2}\right)
^{\alpha /2}\Psi +\left\vert \Psi \right\vert ^{2}\Psi -\left\vert \Psi
\right\vert ^{4}\Psi =0.  \label{NLFSE2}
\end{equation}%
Obviously, when $\alpha =2$ Eq. (\ref{NLFSE2}) reduces to the conventional
cubic-quintic NLSE \cite{CQ12}.

In this paper, we aim to construct spatial vortex-soliton solutions to Eq. (%
\ref{NLFSE2}), with propagation constant $\beta $, as
\begin{equation}
\Psi \left( \zeta ,\xi ,\eta \right) =\psi \left( \xi ,\eta \right)
e^{i\beta \zeta },  \label{Solu}
\end{equation}%
where complex function $\psi (\xi ,\eta )$ obeys a stationary equation:%
\begin{equation}
-\left( -\nabla _{\perp }^{2}\right) ^{\alpha /2}\psi +\left\vert \psi
\right\vert ^{2}\psi -\left\vert \psi \right\vert ^{4}\psi -\beta \psi =0.
\label{NLFSE3}
\end{equation}%
Further, the complex function is expressed in the Madelung's form, $\psi
(\xi ,\eta )=U(\xi ,\eta )e^{i\phi (\xi ,\eta )}$, where real functions $%
U(\xi ,\eta )$ and $\phi (\xi ,\eta )$ represent the amplitude and phase of
the solution. In polar coordinates $(r,\theta )$, related to the underlying
Cartesian coordinates by the usual formulas, $\xi =r\cos \theta $, $\eta
=r\sin \theta $, VSs are expressed as
\begin{equation}
\psi (r,\theta )=U(r)e^{is\theta },  \label{psiU}
\end{equation}%
where positive integer $s$ represents the vorticity (topological charge of
the phase screw dislocation), and the optical intensity\ vanishes as $%
\left\vert A\right\vert ^{2}\sim r^{2s}$ at $r\rightarrow 0$. The power
(integral norm) and total angular momentum of the vortex are%
\begin{equation}
P=\int \int \left\vert \psi \left( \xi ,\eta \right) \right\vert ^{2}d\xi
d\eta ,  \label{Energy}
\end{equation}%
\begin{equation}
L=\int \int \frac{\partial \phi (\xi ,\eta )}{\partial \theta }\left\vert
\psi \left( \xi ,\eta \right) \right\vert ^{2}d\xi d\eta .  \label{AM}
\end{equation}%
As it follows from Eqs. (\ref{psiU})-(\ref{AM}), for stationary VSs the
angular momentum and power are related by $L=sP$.

\section{Numerical results}

\label{Sec III}

\subsection{The numerical method}

\label{Sec IIIa}

The fractional Laplacian in the FSE is defined as a pseudo-differential
operator \cite{JMP51-062102,JMP54-012111,CMA71},%
\begin{equation}
\mathcal{F}[\left( -\nabla _{\perp }^{2}\right) ^{\alpha /2}\psi \left( \xi
,\eta \right) ]=\left( k_{\xi }^{2}+k_{\eta }^{2}\right) ^{\alpha /2}\hat{%
\psi}(k_{\xi },k_{\eta }),  \label{PseudoDO}
\end{equation}%
where $\mathcal{F}$ is the operator of the Fourier transform, and $\hat{\psi}%
(k_{\xi },k_{\eta })$ is the Fourier image of $\psi (\xi ,\eta )$. Below, we
construct vortex-soliton solutions of Eq. (\ref{NLFSE3}) with fractional
values of L\'{e}vy index $\alpha $ by means of the Newton-conjugate-gradient
method \cite{NCG,Book1}. For the sake of comparison, we will also present
usual solutions for $\alpha =2$. Following this method, Eq. (\ref{NLFSE3})
is rewritten in the form of%
\begin{equation}
\Lambda _{0}\psi \left( \xi \right) =0,  \label{NCG1}
\end{equation}%
where the operator is defined as%
\begin{equation}
\Lambda _{0}=-\left( -\nabla _{\perp }^{2}\right) ^{\alpha /2}+\left\vert
\psi \right\vert ^{2}-\left\vert \psi \right\vert ^{4}-\beta .  \label{NCG2}
\end{equation}%
Here, the propagation constant $\beta $ is considered as a given value, and
the solution is calculated by means of Newton iterations,
\begin{equation}
\psi _{n+1}=\psi _{n}+\Delta \psi _{n},  \label{NCG3}
\end{equation}%
where $\psi _{n}$ is an approximate solution, and $\Delta \psi _{n}$ is
computed from the linear Newton-correction equation,%
\begin{equation}
L_{1n}\Delta \psi _{n}=-L_{0}\left( \psi _{n}\right) ,  \label{NCG4}
\end{equation}%
with $L_{1n}$ being the linearization operator of Eq. (\ref{NCG1}),
evaluated with the approximate solution $\psi _{n}$:%
\begin{eqnarray}
L_{1} &=&\left. -\left( -\nabla _{\perp }^{2}\right) ^{\alpha /2}+\left\vert
\psi \right\vert ^{2}-\left\vert \psi \right\vert ^{4}-\beta +\right.
\label{NCG5} \\
&&\left. 2\psi \mathtt{Re}\left( \psi ^{\ast }\right) -4\left\vert \psi
\right\vert ^{2}\psi \mathtt{Re}\left( \psi ^{\ast }\right) .\right.
\end{eqnarray}%
Then, Eq. (\ref{NCG4}) can be solved directly by dint of preconditioned
conjugate gradient iterations \cite{NCG,Book1}.

\subsection{Vortex-soliton (VS)\ solutions}

\label{Sec IIIb}

\begin{figure}[tbp]
\centering\vspace{0cm} \includegraphics[width=12cm]{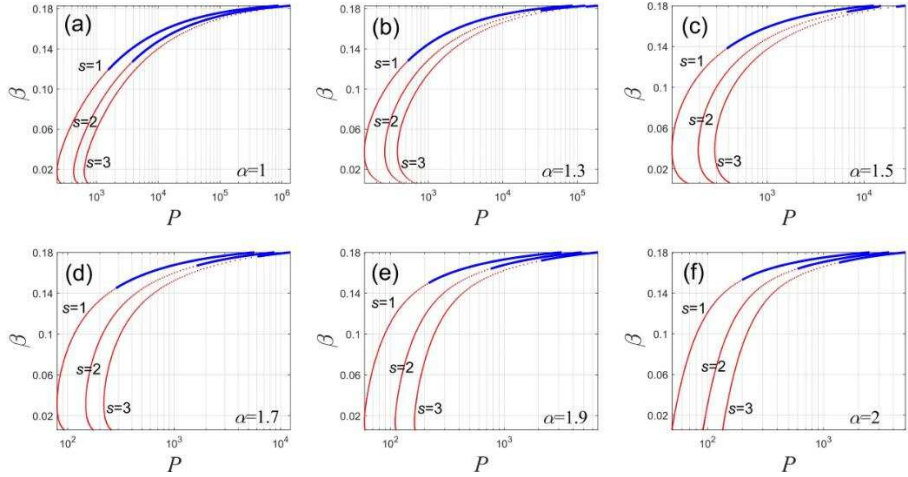}
\vspace{0.0cm}
\caption{(Color online) The propagation constant of vortex solitons versus
their integral power, for vorticities $s=1,2,3$, and different values of the
L\'{e}vy index: (a) $\protect\alpha =1$, (b) $\protect\alpha =1.3$, (c) $%
\protect\alpha =1.5$, (d) $\protect\alpha =1.7$, (e) $\protect\alpha =1.9$,
and (f) $\protect\alpha =2 $. The latter case, corresponding to the usual
NLSE, is included for comparison with the results produced by the fractional
equation. Here, solid blue and dotted red curves denote stable and unstable
solutions, respectively. The limit value of the upper branches at $%
P\rightarrow \infty $ is $\protect\beta _{\max }=3/16$, see the text.}
\label{figure1}
\end{figure}

To quantify numerically found families of VS solutions with different
vorticities $s$ and different values of the L\'{e}vy index $\alpha $, we
display respective dependencies of their propagation constant $\beta $ on
power $P$ in Fig. \ref{figure1} (stability and instability of the families
represented by the solid blue and dotted red lines, respectively, is
identified below by the linear-stability analysis and direct simulations).
It is seen that the celebrated Vakhitov-Kolokolov criterion ($d\beta /dP>0$)
\cite{VK,Fibich} is only a necessary but not sufficient condition for the
stability of the VS solutions belonging to the upper branches in Figs. \ref%
{figure1}(a-d). The limit value of the power for stable (upper) branches of
the $\beta (P)$ dependencies is $\beta _{\max }=3/16$, which is a known
feature of the CQ nonlinearity, that does not depend on the dimension \cite%
{CQ1}-\cite{CQ12}. In the limit of $\beta =\beta _{\max }$, the soliton (in
any dimension) carries over into an exact 1D front solution (alias domain
wall), which connects zero intensity, $\left\vert \psi \right\vert ^{2}=0$,
and the largest value achievable in the soliton, $\left( \left\vert \psi
\right\vert ^{2}\right) _{\max }\equiv \left\vert \psi \left( \beta
=3/16\right) \right\vert ^{2}=3/4$ \cite{Birnbaum}:%
\begin{equation}
\Psi \left( \xi ,\zeta \right) =\exp \left( \frac{3}{16}i\zeta \right) \sqrt{%
\frac{3/4}{1+\exp \left( \pm \sqrt{3}\xi /2\right) }}.  \label{front}
\end{equation}

\begin{figure}[tbp]
\centering\vspace{0cm} \includegraphics[width=9cm]{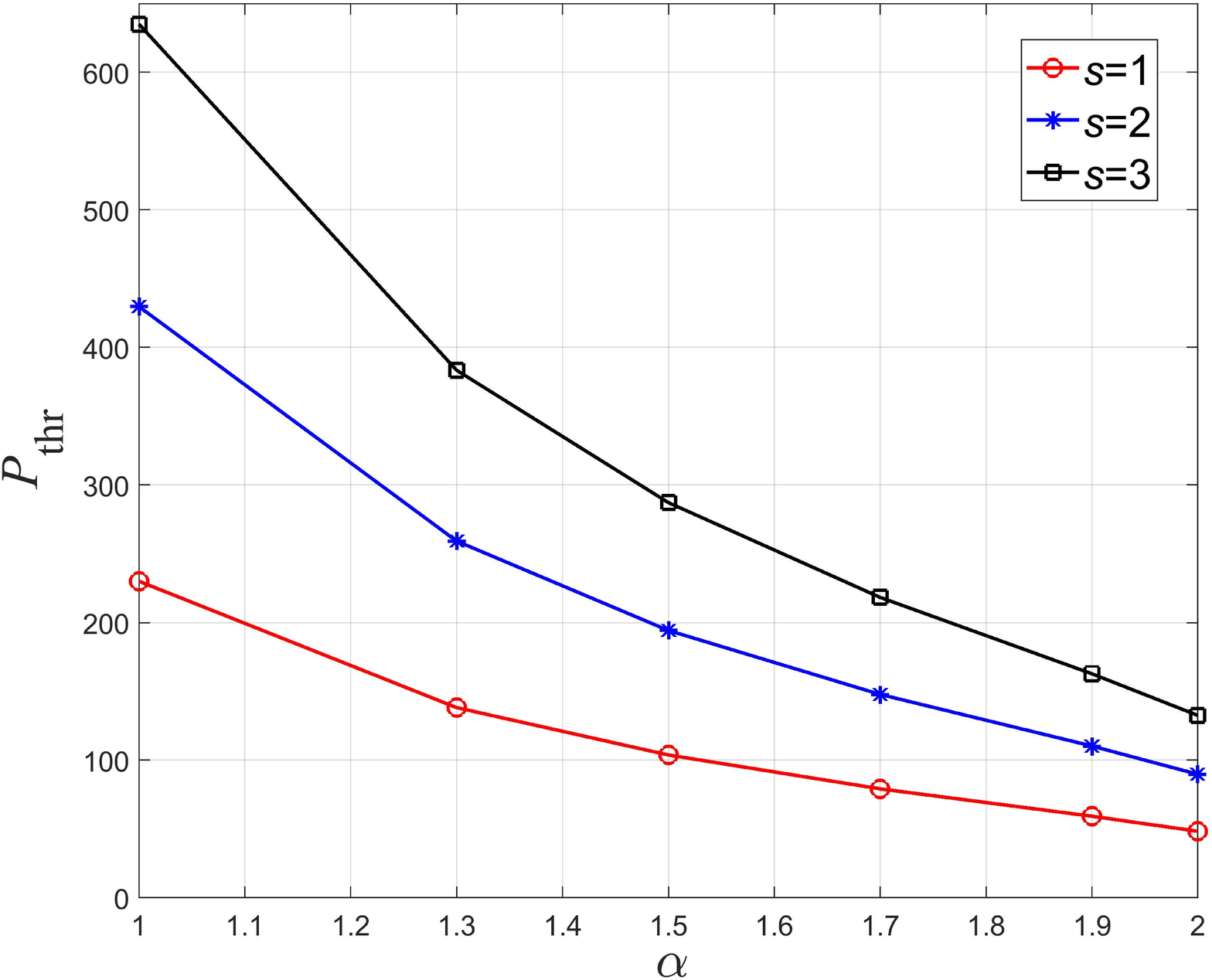}
\vspace{0.0cm}
\caption{(Color online) Dependence of $P_{\mathrm{thr}}$ on the L\'{e}vy
index $\protect\alpha $ of the vortex-soliton solutions for vorticities $%
s=1,2,3$.}
\label{figure2}
\end{figure}

The comparison of panels (a-e) and (f) in Fig. \ref{figure1} demonstrates
that, at all values of the L\'{e}vy index $\alpha <2$, there is a threshold
(minimum) value of the total power, $P_{\mathrm{thr}}$, necessary for the
existence of the VS solutions and, accordingly, there are two branches of
the $\beta (P)$ curves, which merge at $P=P_{\mathrm{thr}}$. In fact, this
feature is similar to one demonstrated by the CQ model in the 3D case \cite%
{CQ2a,CQ2b,CQ8,CQ9}, and the nonexistence of solitons at $P<P_{\mathrm{thr}}$
is explained by the fact the weak nonlinearity cannot balance the
fractional-order diffraction. On the contrary, the threshold vanishes in the
limit of $\alpha =2$, which corresponds to the usual CQ model in 2D.
Dependence of the threshold value of the total power $P_{\mathrm{thr}}$ on
the L\'{e}vy index $\alpha $ is displayed in Fig. \ref{figure2}.

\begin{figure}[tbp]
\centering\vspace{0cm} \includegraphics[width=12cm]{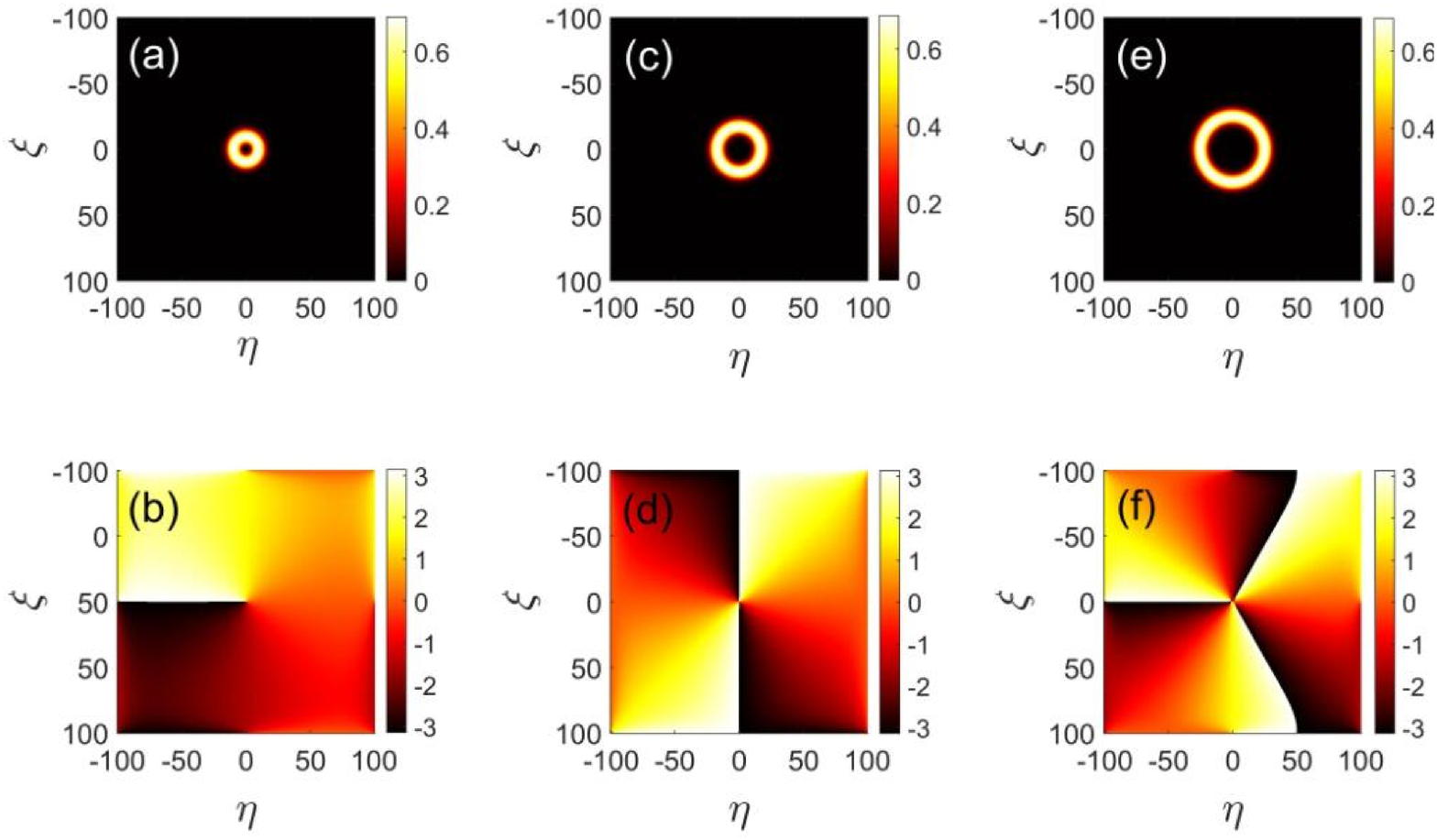}
\vspace{0.0cm}
\caption{(Color online) Intensity and phase patterns of the vortex-soliton
solutions with propagation constant $\protect\beta =0.14$, L\'{e}vy index $%
\protect\alpha =1.5$, and vorticities (a,b) $s=1$ (the 2D soliton is
stable), (c,d) $s=2$ (unstable), (e,f) $s=3$ (unstable).}
\label{figure3}
\end{figure}

\begin{figure}[tbp]
\centering\vspace{-0cm} \includegraphics[width=14cm]{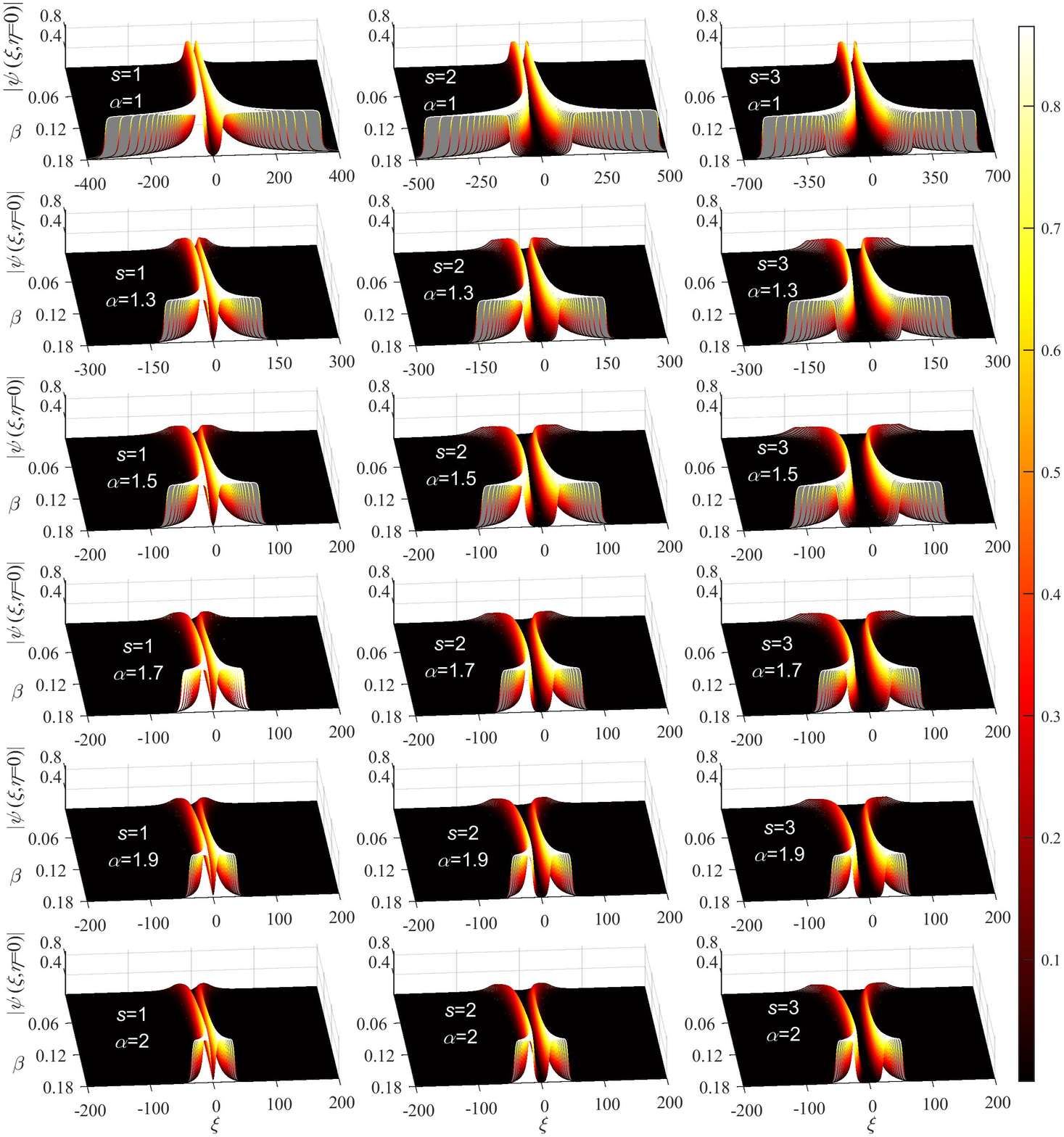}
\vspace{0cm}
\caption{(Color online) Profiles of the cross-section, $\left\vert \protect%
\psi \left( \protect\xi ,\protect\eta =0\right) \right\vert $, of the vortex
solitons for the propagation constant taking values $0.006\leq \protect\beta %
\leq 0.18$ with an interval of $0.001$, and a set of different values of the
vorticity and L\'{e}vy index, from $\protect\alpha =1.0$ to $\protect\alpha %
=2.0$, as indicated in the panels.}
\label{figure4}
\end{figure}

Typical examples of intensity and phase profiles of the numerically
generated VS solutions, with $s=1$, $2$ and $3$, are displayed in Fig. \ref%
{figure3} for $\alpha =1.5$. Further, cross-section profiles of the VS
modes, in the form of $\left\vert \psi \left( \xi ,\eta =0\right)
\right\vert $, are displayed in Fig. \ref{figure4}, for propagation
constants $\beta $ ranging from $0.006$ to $0.18$, which demonstrates the
well-known property of systems with competing nonlinearities, \textit{viz}.,
a trend to the formation of flat-top shapes of the solitons in the limit of
large powers (i.e., $\beta \rightarrow \beta _{\max }=3/16$, $|\psi
|^{2}\rightarrow 3/4$). Further, Fig. \ref{figure4} demonstrates that the
solitons' profiles additionally expand with the decrease of the L\'{e}vy
index, $\alpha $. This trend is explained by the fact that decrease of $%
\alpha $ makes it more difficult for maintaining the balance between the
front layer [see Eq. (\ref{front})], which separates the nearly-constant
value of the intensity inside the soliton, close to $\left( \left\vert \psi
\right\vert ^{2}\right) _{\max }$, and zero intensity outside of the
soliton, and inner pressure of the flat-top soliton.

\subsection{The linear-stability analysis and dynamics}

\label{Sec IIIc}

\begin{figure}[tbp]
\centering\vspace{0cm} \includegraphics[width=14cm]{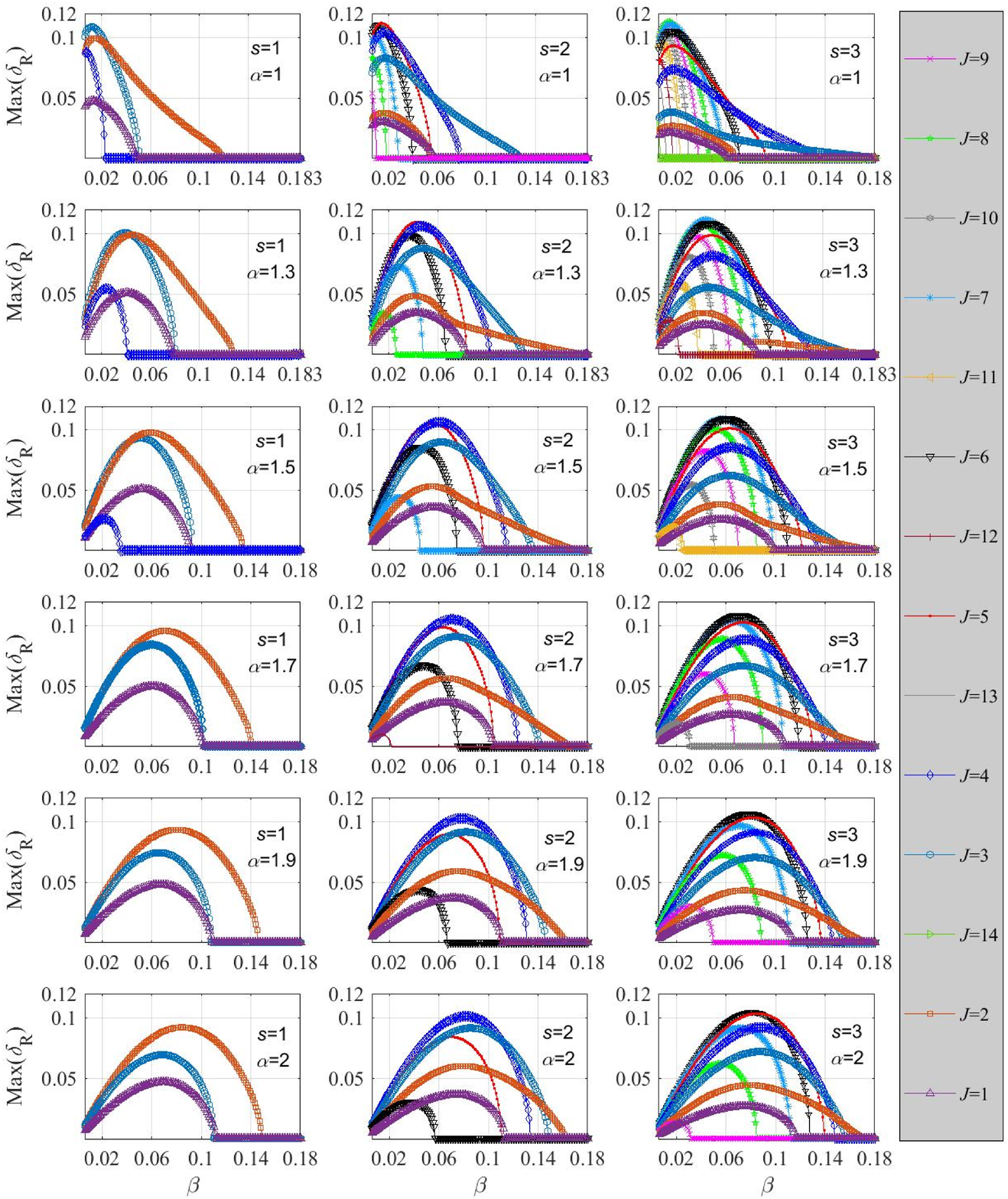}
\vspace{-0.5cm}
\caption{(Color online) The growth rate of perturbation eigenmodes with
azimuthal indices $J$ versus the propagation constant $\protect\beta $ of
the underlying unperturbed vortex soliton, for different values of the L\'{e}%
vy index, from $\protect\alpha =1.0$ to $\protect\alpha =2.0$, as indicated
in the panels.}
\label{figure5}
\end{figure}

As mentioned above, stability of VSs against the splitting instability is a
crucially important issue. First, we address it in the framework of the
linearization of Eq. (\ref{NLFSE2}) for small perturbations (i.e., the
respective Bogoliubov-de Gennes equations), taking the perturbed solution as%
\begin{equation}
\Psi \left( \xi ,\eta ,\zeta \right) =e^{i\beta \zeta }\left[ \psi \left(
\xi ,\eta \right) +\epsilon u\left( \xi ,\eta \right) e^{\delta \zeta
}+\epsilon v^{\ast }\left( \xi ,\eta \right) e^{\delta ^{\ast }\zeta }\right]
,  \label{Perturbation}
\end{equation}%
where $\psi $ is the stationary VS solutions with propagation constant $%
\beta $, taken as per Eq. (\ref{psiU}), while $u\left( \xi ,\eta \right) $
and $v\left( \xi ,\eta \right) $ represent eigenmodes of perturbations, with
an infinitesimal amplitude $\epsilon $. In cylindrical coordinates $%
(r,\theta ,\zeta )$, expression (\ref{Perturbation}), including disturbances
characterized by an integer azimuthal index, $J$, and instability growth
rate, $\delta $ (which may be complex), is written as
\begin{eqnarray}
\Psi \left( r,\theta ,\zeta \right)  &=&e^{i\left( \beta \zeta +s\theta
\right) }\left[ U\left( r\right) +\epsilon \varepsilon _{J}^{+}\left(
r\right) e^{+iJ\theta }e^{\delta \zeta }\right.   \label{PerturbationC} \\
&&\left. +\epsilon \varepsilon _{J}^{-}\left( r\right) e^{-iJ\theta
}e^{\delta ^{\ast }\zeta }\right] ,
\end{eqnarray}%
where $\ast $ stands for the complex conjugate, while $\varepsilon
_{J}^{+}\left( r\right) $\ and $\varepsilon _{J}^{-}\left( r\right) $\ are
the respective perturbation eigenmodes.

Substituting Eq. (\ref{Perturbation}) into Eq. (\ref{NLFSE2}) and the
linearization lead to the following eigenvalue problem:%
\begin{equation}
i\left(
\begin{array}{cc}
\mathcal{L}_{11} & \mathcal{L}_{12} \\
\mathcal{L}_{21} & \mathcal{L}_{22}%
\end{array}%
\right) \left(
\begin{array}{c}
u \\
v%
\end{array}%
\right) =\delta \left(
\begin{array}{c}
u \\
v%
\end{array}%
\right) ,  \label{LinearEigEqs}
\end{equation}%
with matrix elements%
\begin{equation}
\mathcal{L}_{11}=+\left[ -\left( -\nabla _{\perp }^{2}\right) ^{\alpha
/2}+2\left\vert \psi \right\vert ^{2}-3\left\vert \psi \right\vert
^{4}-\beta \right] ,  \label{L11}
\end{equation}%
\begin{equation}
\mathcal{L}_{12}=+\psi ^{2}+2\left\vert \psi \right\vert ^{2}\psi ^{2},
\label{L12}
\end{equation}%
\begin{equation}
\mathcal{L}_{21}=-\psi ^{\ast 2}-2\left\vert \psi \right\vert ^{2}\psi
^{\ast 2},  \label{L21}
\end{equation}%
\begin{equation}
\mathcal{L}_{22}=-\left[ -\left( -\nabla _{\perp }^{2}\right) ^{\alpha
/2}+2\left\vert \psi \right\vert ^{2}-3\left\vert \psi \right\vert
^{4}-\beta \right]  \label{L22}
\end{equation}%
where $\delta $ is a complex eigenvalue of Eq. (\ref{LinearEigEqs}).
Obviously, stability demands to have only pure imaginary eigenvalues $\mu $.

\begin{figure}[tbp]
\centering\vspace{0cm} \includegraphics[width=12cm]{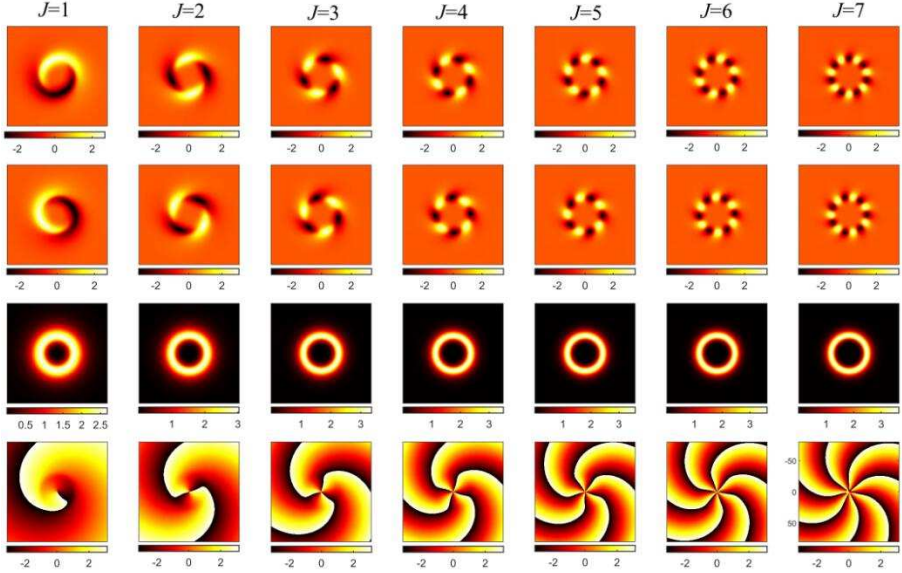} \vspace{0cm}
\vspace{-0cm}
\caption{(Color online) Perturbation eigenmodes for the vortex soliton with $%
s=2$ and $\protect\beta=2$, with L\'{e}vy index $\protect\alpha =1.5$ and
different values of the eigenmode's azimuthal index, $J$. The first and
second rows depict the real and imaginary parts of the eigenmodes, while the
third and fourth rows show their intensity and phase patterns. In all
panels, the displayed domain is $[-80,80]\times \lbrack -80,80]$.}
\label{figure6}
\end{figure}

The linear problem based on Eq. (\ref{LinearEigEqs}) can be solved by means
of the Newton-conjugate-gradient method \cite{Book1} (written in the
Cartesian coordinates), with the initial guess for the perturbation
eigenmodes taken as per Eq. (\ref{PerturbationC}). Dependencies of the so
found largest perturbation growth rates, $\delta _{\mathrm{R}}\equiv $Re$%
(\delta )$, on the propagation constant, for different values of $s$, $J$,
and $\alpha $, are summarized in Fig. \ref{figure5}. It is seen that, for
the unitary VS ($s=1$) at $\alpha =1.0$, $\alpha =1.3$, and $\alpha =1.5$,
the azimuthal index of the dominant perturbation eigenmode switches from $J=3
$ to $J=2$ with the increase of $\beta $, which is different from the case
of the usual NLSE with $\alpha =2$. The decrease of the L\'{e}vy index also
leads to switch of the azimuthal index of the dominant perturbation
eigenmode for the VS solutions with $s=2$ and $3$. Especially, for the VS
solutions with $s=2$ at $\alpha =1.0$, the perturbation eigenmode with the
azimuthal index $J=2$ (it is one of the dominant perturbation eigenmodes for
the VS solutions with $s=2$ ranging from $\alpha =1.3$ to $\alpha =2$) is
suppressed, but, the azimuthal index $J=3$ becomes the dominant perturbation
eigenmode with the increase of $\beta $. This lead to the stability area of
the VSs with $s=2$ sharply shrinks for $\alpha=1$. Typical examples of
perturbation eigenmodes $\varepsilon _{J}^{+}$\ and $\varepsilon _{J}^{-}$\
for the VSs with different values of $J$, are displayed in Fig. \ref{figure6}%
.

The impact of the L\'{e}vy index on the stability area of the VSs, in which
all eigenvalues $\delta $ are imaginary, is summarized in Table \ref{table1}%
, while ${\beta }_{\,\min }{(s=3)}$ corresponds to the VSs with radii so
large that it is difficult to find those values with sufficient accuracy.

To verify the validity of the results, the stability boundaries of the VS
solutions in the usual NLSE with $\alpha =2$ was also recalculated by means
of the numerical method adopted in this work, with a conclusion that the
results are identical to those previously reported in Refs. \cite{CQ6b,CQ12}%
. A noteworthy feature revealed by Table \ref{table1} is that the stability
region of the unitary VS with $s=1$ shrinks with the decrease of the L\'{e}%
vy index from $\alpha =1$ up to $\alpha =2$, while, on the contrary, it
expands for the VSs with $3$. It is worth noting that the stability area of
the VSs with $s=2$ expands for $1<\alpha \leq 2$, but sharply shrinks for $%
\alpha=1$.

\begin{table*}[tbph]
\centering%
\resizebox{\textwidth}{!}{
\begin{tabular}{ccccccccc}
\hline\hline
&$s$ && $\alpha =1$ & $\alpha =1.3$ & $\alpha =1.5$ & $\alpha =1.7$ & $\alpha =1.9$ & $\alpha =2$ \\ \hline
&$1$ && $0.119(1553)$ & $0.128(532)$ & $0.135(383)$ & $0.141(285)$ & $0.147(217)$ & $0.149(199) $ \\ \hline
&$2$ && $0.127(3822)$ & $0.178(32385)$ & $0.170(6651)$ & $0.165(1632)$ & $0.163(761)$ & $0.162(600)$ \\ \hline
&$3$ &&  & $0.182(130586)$ & $0.179(21546)$ & $0.176(6112)$ & $0.173(2113)$ & $0.170(1364)$ \\ \hline\hline
\end{tabular}}%
\caption{Vortex solitons are stable in intervals of the propagation constant
$\protect\beta _{\min }\leq \protect\beta <\protect\beta _{\max }\equiv 3/16$%
, with the lower boundary $\protect\beta _{\min }$($P_{\min }$) shown in
entries of the table, where $P_{\min }$ is the corresponding value of the
power, for different values of vorticity $s$ and L\'{e}vy index $\protect%
\alpha $.}
\label{table1}
\end{table*}

\begin{figure}[tbp]
\centering\vspace{0cm} \includegraphics[width=12cm]{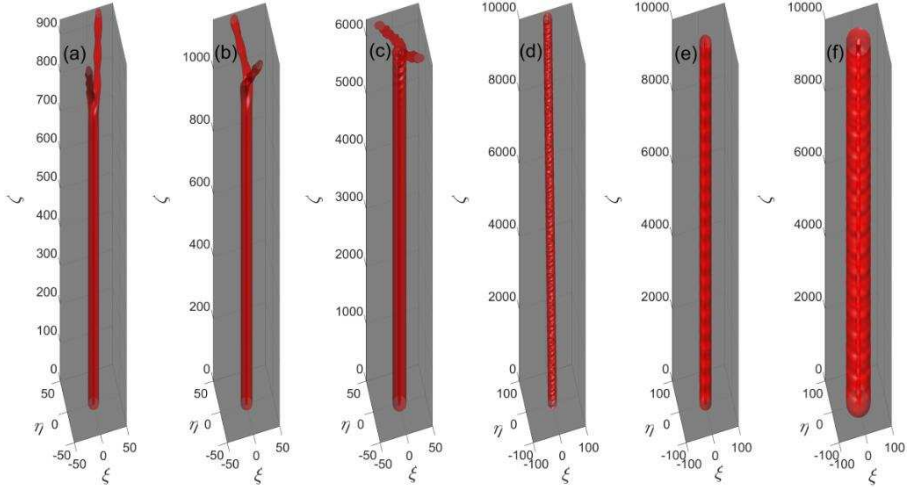}
\vspace{0.0cm}
\caption{(Color online) Examples of the simulated evolution of the unitary
vortex solitons with $s=1$ and propagation constant $\protect\beta =0.14$,
for different values of the L\'{e}vy index, $\protect\alpha $. The evolution
is displayed by isosurfaces drawn at the local intensity, $\left\vert \Psi
\right\vert ^{2}$, fixed at the half-maximum level. The evolution is
unstable for $\protect\alpha =2.0$ (a), $\protect\alpha =1.9$ (b), and $%
\protect\alpha =1.7$ (c). In panels (d), (e), and (f) stable evolutions are
shown for $\protect\alpha =1.5$, $\protect\alpha =1.3$, and $\protect\alpha %
=1.0$ under the action of random-noise perturbation at the $5\%$ amplitude
level.}
\label{figure7}
\end{figure}

\begin{figure}[tbp]
\centering\vspace{0.1cm} \includegraphics[width=12cm]{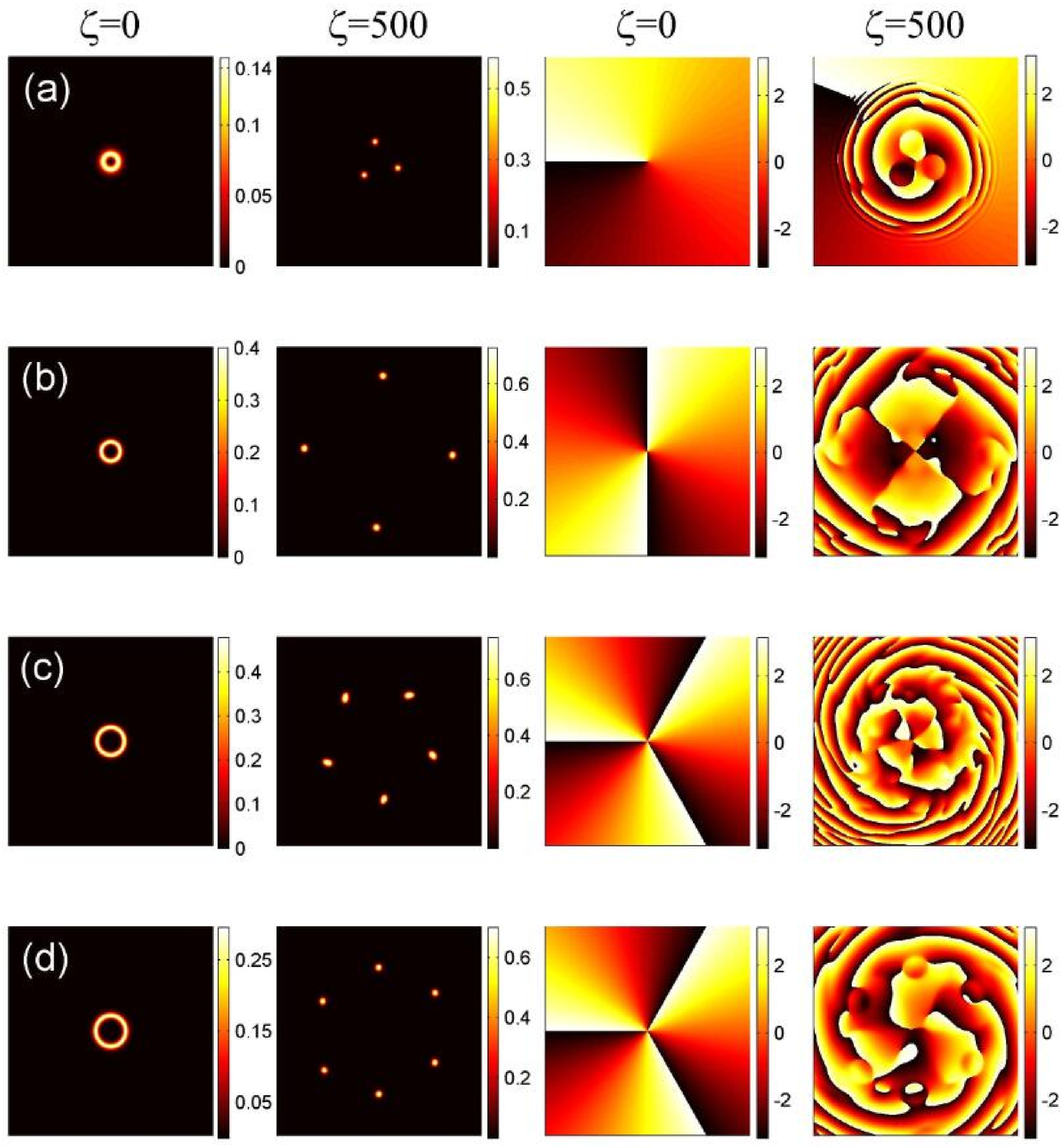}
\vspace{0.0cm}
\caption{(Color online) Splitting of unstable vortex-soliton solutions at $%
\protect\alpha =1.5$. The first and second columns show intensities of the
input and output, while the third and last ones show the same for the
corresponding phases. (a) The unitary vortex soliton, with $s=1$ and $%
\protect\beta =0.03 $. (b) The soliton with $s=2$ and $\protect\beta = 0.08$%
. (c) and (d): The solitons with $s=3$ and $\protect\beta =$ $0.095$ and $%
\protect\beta =0.06$, respectively. All panels are displayed in the domain
of size $[-150,150]\times \lbrack -150,150]$.}
\label{figure8}
\end{figure}

Results predicted by the linear-stability analysis and collected in Fig. \ref%
{figure5} and Table \ref{table1} were verified by direct simulations of the
perturbed simulations of the VSs, performed in the framework of Eq. (\ref%
{NLFSE2}). First, the (in)stability of the unitary VS solutions with $s=1$
and fixed $\beta =0.14$ is tested for different values of the L\'{e}vy
index. The simulations, displayed in Fig. \ref{figure7}, indicate that, in
agreement with the predictions of the linear-stability analysis, in panels
(a-c) the solutions are unstable against azimuthal perturbations, splitting
into two fragments when the L\'{e}vy index takes values $\alpha =2$, $1.9$
and $1.7$. Further, the distance of stable propagation increases with the
decrease of the L\'{e}vy index. Namely, at $\alpha =2$ ($\delta _{R}\approx
0.0373$), the splitting starts relatively fast in Fig. \ref{figure7}(a), at $%
\zeta \approx 800$, while, at $\alpha =1.9$ ($\delta _{R}\approx 0.0284$),
it starts at $\zeta \approx 1000$ in Fig. \ref{figure7}(b). Further, at $%
\alpha =1.7$ the instability growth rate is very small, $\delta _{R}\approx
0.0041$, and, accordingly, the VS breaks only at $\zeta \approx 6100$ in
Fig. \ref{figure7}(c). Finally, also in agreement with the linear-stability
prediction, at $\alpha =1.5$, $\alpha =1.3$, and $\alpha =1.0$ the
evolutions of the VSs are completely stable in Figs. \ref{figure7}(d), \ref%
{figure7}(e), and \ref{figure7}(f), even if it is perturbed by azimuthal
random noise at the $5\%$ amplitude level, and the simulation is extremely
long: the total propagation length, $\zeta =10000$, roughly corresponds to $%
100$ characteristic diffraction lengths of the beam. This length is
estimated, in term of the beam's width $W$, as $\zeta _{\mathrm{diffr}}\sim
W^{\alpha }$. Additional long-scale simulations of the evolution of stable
VSs under the action of noisy perturbations demonstrates slow random drift
of the soliton's center, and, in some cases, small deformation of the
intensity ring. The increase of $\beta $, i.e., as a matter of fact, of the
soliton's total power, naturally leads to suppression of the drift and
deformation. This dynamics is illustrated by movies in Supplemental Material
\cite{SM1}.

Finally, in Fig. \ref{figure8} we display additional results illustrating
the development of splitting of unstable VSs, with vorticities $s=1,2$ and $%
3 $. In all cases, it is concluded that the number of fragments (whose
number grows from three to six with the increase of $s$) is determined by
the fastest growing eigenmode of the azimuthal perturbations, in full
agreement with the linear-stability analysis. Further details of the
splitting dynamics are presented in movies included in Supplemental Material
\cite{SM2}.

\section{Conclusion}

\label{Sec IV}

We have numerically investigated the existence and stability of families of
spatial VS (vortex-soliton) solutions in the fractional NLSE with competing
CQ (cubic-quintic) nonlinearities. A crucial difference from the usual
two-dimensional NLSE, where there is no threshold for the existence of VS
solutions, is that in the fractional model they exist at values of the
soliton's total power, $P$, exceeding a finite threshold. Accordingly, the
dependence of the propagation constant on $P$ consists of two branches,
which merge at $P=P_{\mathrm{thr}}$. Detailed results are reported for
vorticities $s=1,2,$ and $3$.

The stability of the VS solutions has been investigated by means of the
linearized equation for small perturbations. A remarkable conclusion is that
the decrease of the L\'{e}vy index of the fractional NLSE leads to shrinkage
of the stability area of the unitary ($s=1$) VS solutions, on the contrary,
it expands for the solitons with $3$, especially, the stability area of the
VSs with $s=2$ expands for $1<\alpha \leq 2$ but shrinks for $\alpha=1$. The
predictions of the linear-stability analysis are fully corroborated by
direct simulations, which, in particular, demonstrate splitting of unstable
VSs in sets of fragments the number of which is exactly predicted by the
dominant eigenmode of small perturbations.

As an extension of the present work, it may be interesting to address motion
and collisions of stable VSs (note that the fractional NLSE\ is not a
Galilean invariant one, therefore the mobility of solitons is a nontrivial
issue).

\section{Acknowledgments}

This work was supported by the National Natural Science Foundation of China
(NNSFC) (11805141, 11804246) and Shanxi Province Science Foundation for
Youths (201901D211424), and supported by Scientific and Technological
Innovation Programs of Higher Education Institutions in Shanxi (STIP)
(2019L0782). The work of BAM was supported, in a part, by the Israel Science
Foundation through grant No. 1286/17.

\end{document}